# Optical Convolutional Spectrometer

Chunhui Yao[1,2]†, Jie Ma[2]†, Ningning Wang[2], Peng Bao[1], Wei Zhuo[2], Tao Zhang[2], Wanlu Zhang[1], Kangning Xu[2], Ting Yan[2], Liang Ming[2], Yuxiao Ye[2], Tawfique Hasan[1], Ian White[1,3], Richard Penty[1], and Qixiang Cheng[1,2]*

1. Department of Engineering, University of Cambridge, UK
2. GlitterinTech Limited, Xuzhou, China
3. University of Bath, Bath, UK
*qc223@cam.ac.uk;
†These two authors contribute equally to this work.

**Abstract**

Optical spectrometers are fundamental across numerous disciplines in science and technology. However, miniaturized versions, while essential for in situ measurements, are often restricted to coarse identification of signature peaks and inadequate for metrological purposes. Here, we introduce a new class of spectrometer, leveraging the convolution theorem as its mathematical foundation. Our 'convolutional spectrometer' offers unmatched performance for miniaturized systems and distinct structural and computational simplicity, featuring a centimeter-scale footprint for the fully packaged unit, low cost (~$10) and a 2400 cm$^{-1}$ (approximately 500 nm) bandwidth. We achieve excellent precision in resolving complex spectra with sub-second sampling and processing time, demonstrating a wide range of applications from industrial and agricultural analysis to healthcare monitoring. Specifically, our spectrometer system classifies diverse solid samples, including plastics, pharmaceuticals, coffee, flour and tea, with 100% success rate, and quantifies concentrations of aqueous and organic solutions with detection accuracy surpassing commercial benchtop spectrometers. We also realize the non-invasive sensing of human biomarkers, such as skin moisture (mean absolute error; MAE = 2.49%), blood alcohol (1.70 mg dL$^{-1}$), blood lactate (0.81 mmol L$^{-1}$), and blood glucose (0.36 mmol L$^{-1}$), highlighting the potential of this new class of spectrometers for low-cost, high-precision, portable/wearable spectral metrology.

**Main**

The development of miniaturized spectrometers has long been an important pursuit in both academia and industry, enabling spectral analysis beyond conventional laboratory constraints[1–3]. However, their coarse performance falls notably short in meeting today's demands for capturing real-time in situ spectroscopic data through embeddable smart devices[4,5]. For example, infrared spectroscopy typically necessitates fine resolution, high precision, and ultra-wide spectral range to acquire sufficient spectral information[6,7]. Practical use cases also prioritize other criteria, including noise tolerance, sampling speed, processing complexity, and temperature resilience[8,9]. In particular, these rigorous, multifaceted requirements are pronounced for healthcare applications[10–12]. Many biomarkers of interest, such as blood lactate and glucose, exist at low concentrations and share overlapped absorption bands, making their data modeling and feature extraction a formidable task[13–15].

In recent years, the rapid advancement of photonic integration technologies has paved the way for developing next-generation chip-scale spectrometer with mass-manufacturability[16,17]. To date, however, the reported integrated spectrometers are mostly at bare-die level for proof-of-concept demonstration, allowing only basic spectrum recovery[18–20]. In general, the underlying operational principles classify existing spectrometers into dispersive, narrowband filtering, Fourier transform, and computational reconstruction types (see Fig. 1a)[5]. For the first three types, the bandwidth and resolution are inevitably limited by the achievable optical path lengths/differences or the scale of the filter arrays, especially for those implemented in planar photonic integrated circuits (PICs)[21–23]. The reconstructive type leverages compressive sampling schemes to break the performance-size trade-off, though at the cost of laborious calibration processes and substantial computing resources for reconstruction[24–26]. In addition, the non-linearity of reconstructive algorithms makes them highly sensitive to measurement noise, leading to unpredictable distortions in the recovered spectrum[27–29].

Here, we propose a new class of spectrometer by harnessing one of the most fundamental principles in signal processing—the convolution theorem—to retrieve arbitrary incident spectra. It employs a simple cascade of optical components with periodic spectral responses, such as unbalanced Mach-Zehnder interferometers (MZIs) or micro-ring resonators (MRRs), to form an overlaid response with systematic periodicity. By applying proportional phase tuning to individual components, the overlaid system response can be linearly shifted in the spectral domain. This waveform shifting physically executes the circular convolution operation on the input spectrum, enabling its recovery via (inverse) Fourier transforms. Such a linear convolutional scheme, owing to its underlying mathematical nature, offers negligible computation load, a relaxed calibration process, and strong noise tolerance. Moreover, the system's inherent periodicity permits the circular convolution to occur within any of its cycles, theoretically granting an unlimited operational bandwidth. Its resolution can also be exponentially escalated by increasing the number of cascading components and tailoring their individual spectral responses. These attractive features distinguish our convolutional spectrometer (ConvSpec) from all existing types (see Table 1 in Discussion and Conclusion).

Experimentally, we implement a ConvSpec on a SiN integration platform and encapsulate it within a comprehensive optoelectronic package, achieving a centimeter-scale overall footprint, a < 0.4 second sampling and local processing time, and a total cost of around \$10. Meanwhile, our device operates over an ultra-wide spectral range from 5900 $cm^{-1}$ to 8300 $cm^{-1}$ (i.e. about 1200 nm to 1700 nm), with exceptional accuracy even in resolving complex spectra. Its thermal robustness is also validated by withstanding an extreme temperature variation from −20°C to 80°C.

Our ConvSpec enables real-time, high-precision, and data-intensive metrological measurements across various near-infrared (NIR) sensing applications. We first demonstrate the classification of solid substances, such as different types, origins, or grades of plastics, pharmaceuticals, coffee, flour, and tea, all attaining a 100% success rate. It also quantifies the concentration/purity of diverse aqueous and organic solutions with a detection accuracy surpassing commercial benchtop spectrometers. Furthermore, our device demonstrates non-invasive sensing of human biomarkers under intricate physiological conditions. It not only realizes the quantitative assessment of skin moisture, but also accurately measures low-concentration subsurface biomolecules, including blood alcohol, blood lactate, and blood glucose. Additionally, it achieves the single-participant long-term tracking of daily glucose fluctuations. This benchmarks the transformation of miniaturized spectrometers into embeddable spectrometric sensors, empowering a broad range of applications from material analysis to healthcare solutions.

## Results

### Principles and Theoretical Analysis

Figure 1a outlines the underlying principles of current spectrometer types, including spectral demultiplexing with linear readout, interferometry with Fourier transform, and sampling matrices with algorithmic reconstruction, together with the proposed ConvSpec that leverages the circular convolution in the spectral domain. Incorporating only a few optical components with periodic responses, our ConvSpec can be readily implemented on various material platforms using different building blocks, such as MZIs, MRRs, Bragg gratings and Fabry-Pérot cavities (see discussions in Supplementary

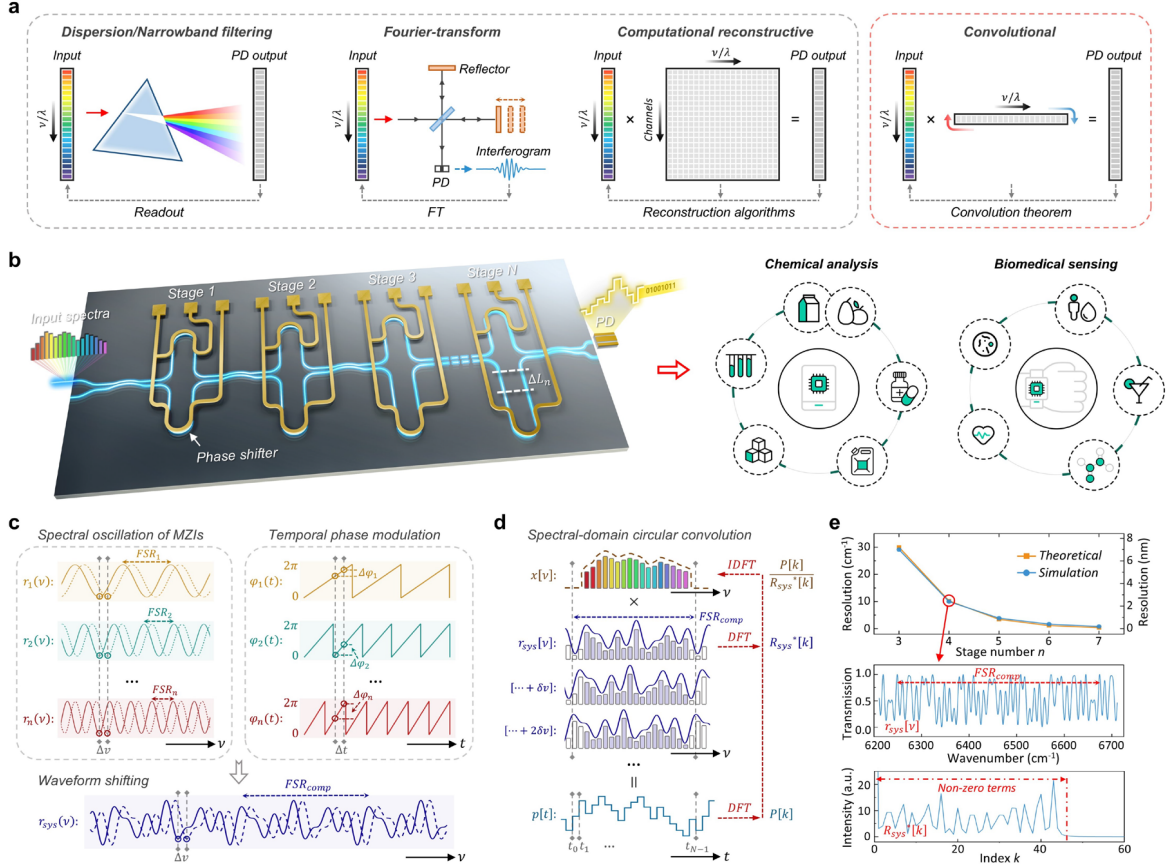

**Fig. 1 | Defining a new class of spectrometer based on the convolution theorem.** (a) Working principles of various spectrometer categories, compared with the proposed convolutional spectrometer. PD: photodetector. (b) Schematic (left) of a convolutional spectrometer with cascaded unbalanced MZIs on a photonic integration platform. The conceptual diagram (right) highlights its potential for portable/wearable devices, accommodating a wide range of spectroscopic applications. (c) Diagrammatic illustration of how applying proportional phase modulation to MZI stages with varying FSRs enables spectral-domain waveform shifting of the overlaid system response. (d) The underlying mechanism of how spectral-domain shifting of the system response executes the circular convolution operation on an incident spectrum, yielding corresponding PD outputs. The right-side arrows illustrate the steps of calculating the convolution theorem for spectrum retrieval. (e) Spectrometer resolution vs. the number of cascading stages (with a fixed composite FSR at 420 cm⁻¹). For example, the inset plots the simulated system response of a 4-stage convolutional core and its corresponding frequency components after DFT.

Information Section 1). Here, we employ a cascade of unbalanced MZIs on a SiN integration platform, as illustrated in Fig. 1b. Denoting the time-varying spectral response of an individual MZI stage $i$ as $r_i(\nu, t)$, where $\nu$ represents the wavenumber, the cascading system exhibits an overlaid system response $r_{sys}(\nu, t)$ that equals the product of all individual stage responses. The free spectral range (FSR) of each MZI is given by $FSR_i = 1/n_g \Delta L_i$, where $n_g$ denotes the group index and $\Delta L_i$ is the arm length difference. Consequently, the cascading system also features a composite FSR, written as:

$$FSR_{comp} = LCM[FSR_i, \ i = 1, 2, \ldots n_{stage}] \tag{1}$$

where $LCM[\ldots]$ represents the least common multiple. Figure 1c illustrates that by applying phase tuning $\varphi_i(t)$ inversely proportional to each MZI's FSR, the overlaid waveform can be circularly shifted over the composite FSR (see derivations in Methods).

Figure 1d further illustrates how such waveform shifting is utilized to execute the circular convolution operation. For an incident spectrum $x(\nu)$ that falls within the span of any composite FSR, the temporal output power $p(t)$ can be expressed as:

$$p(t) = \int_{v_{start}}^{v_{end}} x(v) r_{sys}(v,t) dv \quad (2)$$

where $v_{end} - v_{start} = FSR_{comp}$. Eq. (2) can then be discretized into a sequence format, as:

$$p[t] = \sum_{v=v_0}^{v_{N-1}} x[v] r_{sys}[(v + \delta v \cdot t)_N], \quad t = 0,1,2,\ldots,N-1$$
$$= \sum_{v=v_0}^{v_{N-1}} x[v] r_{sys}{}^*[(-v + \delta v \cdot t)_N], \quad t = 0,1,2,\ldots,N-1 \quad (3)$$

where $\delta v$ represents the smallest footstep in waveform shifting, $t$ is discretized time index, and $(\cdot)_N$ denotes the circular shift of the sequence modulo $N$, which equals the division between $FSR_{comp}$ and $\delta v$. $r_{sys}{}^*[v]$ refers to the flip of the original sequence $r_{sys}[v]$. Notably, Eq. (3) strictly adheres to the circular convolution theorem[30], allowing its transformation via Discrete Fourier Transform (DFT) to yield:

$$P[k] = R_{sys}{}^*[k] X[k], \quad k = 0,1,\ldots,N-1 \quad (4)$$

where $k$ denotes the index of the discretized sequence. Therefore, the incident spectrum can be retrieved through the inverse DFT, as:

$$x[v] = IDFT\left(\frac{P[k]}{R_{sys}{}^*[k]}\right) = IDFT\left(\frac{DFT(p[t])}{DFT(r_{sys}{}^*[v])}\right) \quad (5)$$

Equation (5) reveals the core mathematical mechanism of our ConvSpec, from which two key features can be noted: 1) the periodic nature allows the convolution operation to occur within any composite FSR, while the span of composite FSR can be manipulated by tailoring the FSRs of individual MZIs; 2) as $R_{sys}{}^*[k]$ appears in the denominator, its highest frequency component—i.e. the largest non-zero term in the $R_{sys}{}^*[k]$ sequence—determines the maximum resolvable frequency component in the input spectra, thus defining the spectrometer resolution, as:

$$resolution = FSR_{comp}/max\{k|R_{sys}{}^*[k] > 0\} \quad (6)$$

Hence, it is crucial for the convolutional core to accommodate extensive frequency components, covering a continuous range from low to high frequency regions. This can be achieved by systematically optimizing the spectral properties of each MZI and increasing the number of stages. Rigorous derivations and simulations illustrate that as stage number increases, the frequency components embedded in the system response grow exponentially, enabling an exponential increase in resolution as per Eq. (6) (see Supplementary Information Section 2). For example, Fig. 1e shows that with the composite FSR set at around 420 cm$^{-1}$ (i.e. 100 nm), scaling the number of stages from three to seven enhances the resolution from about 30 cm$^{-1}$ (i.e. 7 nm) to below 0.5 cm$^{-1}$ (i.e. 0.1 nm). The inset showcases the optimized system response for a 4-stage convolutional core, along with its corresponding frequency components.

Besides, the inherent linearity of the convolution operation ensures that any measurement noise is linearly imposed onto the resolved spectra without amplification or distortion. This permits the application of regular noise reduction techniques, such as digital low-pass filters, to effectively enhance the spectrometer accuracy[31]. In practice, the dispersion effects that stretch the FSR periods in the spectral domain can also be compensated by mathematically amending Eq. (5). Detailed noise and dispersion correction schemes are provided in Supplementary Information Section 3.

**Device Design and Characterization**

Figures 2a-b present the overall view of our integrated ConvSpec chip, mounted on a customized printed circuit board (PCB) and edge-coupled with optical fibers (see Methods for details). The PCB integrates an MCU for circuit control and data processing, an InGaAs PD with read-out circuits for optical signal detection, and auxiliary driving circuits for phase modulation. Our device features a compact overall size of 4.8 × 6.2 × 0.6 cm$^3$, a lightweight of < 16.5 g, and $10-level total cost (see breakdowns in Supplementary Table S1). The photonic chip (0.9 × 3.4 mm$^2$) cascades four stages of unbalanced MZIs, offering a balanced trade-off between the resolution, footprint, and control

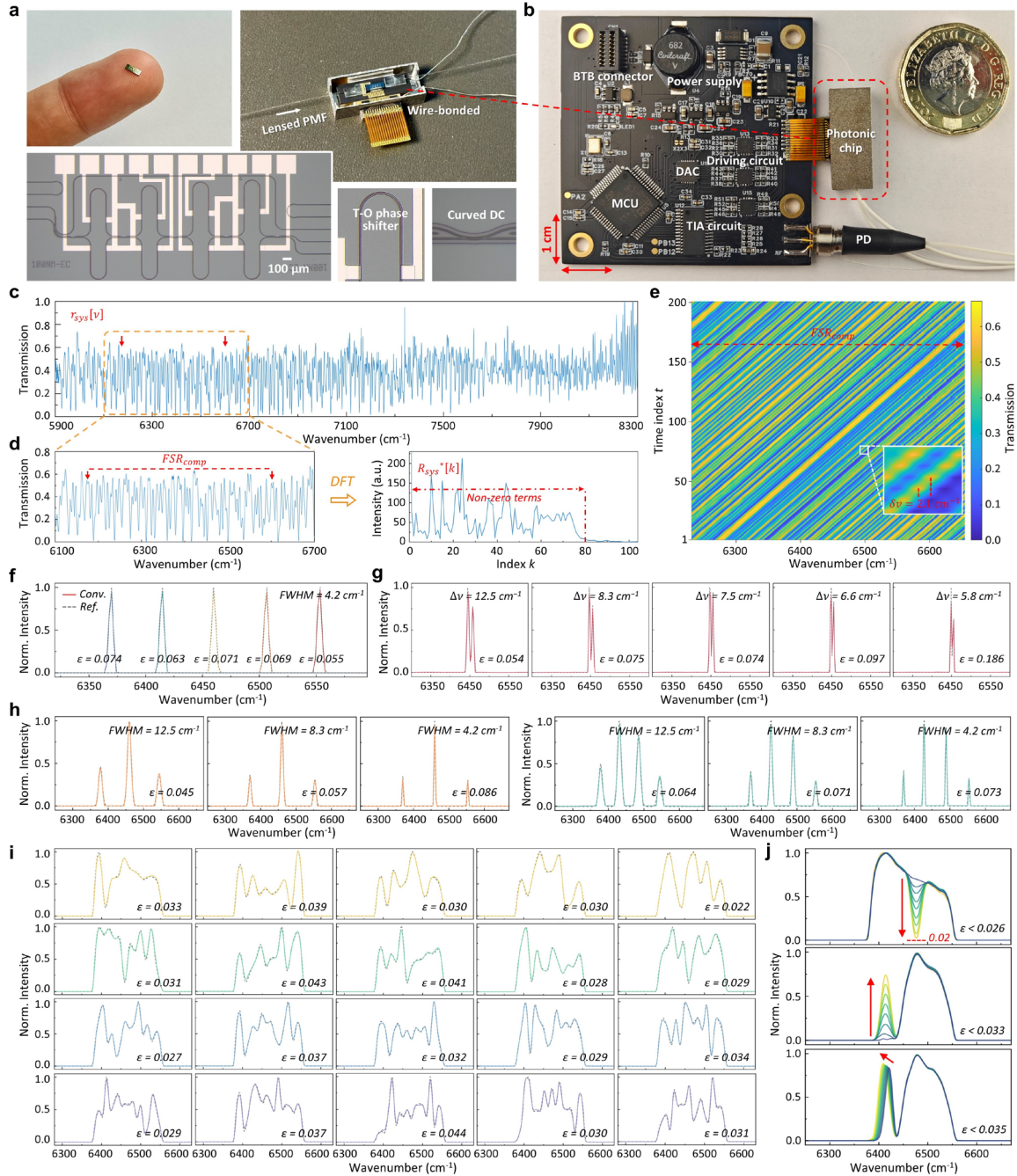

**Fig. 2| Characterization of spectrometer performance.** (a) Photos of the spectrometer chip, wire-bonded and optically coupled with lensed polarization maintaining fiber (PMF). Insets enlarge the thermo-optic (T-O) phase shifter and curved directional coupler (DC), respectively. (b) Photo of the fully packaged ConvSpec with a customized printed circuit board. MCU: microcontroller unit, DAC: digital-to-analog converter, TIA: transimpedance amplifier. (c) Measured system response across an ultra-wide spectral range, exhibiting a systematic periodicity of around 420 cm$^{-1}$. (d) System response within a specific composite FSR and its corresponding frequency components after DFT. The red arrows mark the boundaries of this composite FSR, highlighting the repetition of notable spectral peaks. (e) Measured temporal circular waveform shifting. The inset shows a shifting step of 2.1 cm$^{-1}$. (f) Resolved single-peak spectra centered at varying wavenumbers. (g) Resolved dual-peak spectra with decreasing spectral spacing. (h) Resolved tri- and quad-peak spectra with different FWHMs. (i) Resolved randomly shaped spectra, with waveform complexity increasing progressively from top to bottom row. (j) Resolved spectra with gradually dipping, rising, and shifting peaks, respectively.

complexity. Its integration on a SiN platform ensures minimal loss, low dispersion, and temperature insensitivity. Curved directional couplers (DCs) are employed in the MZIs to facilitate an ultra-wide bandwidth (Supplementary Information Section 5). Figure 2c shows the measured system response over a 2400 cm$^{-1}$ spectral range from 5900 cm$^{-1}$ to 8300 cm$^{-1}$ (i.e. 1200 nm to 1700 nm), exhibiting a complicated waveform but with consistent periodicity of around 420 cm$^{-1}$. Figure 2d further plots the system response within one specific composite FSR and its corresponding frequency components, which contain abundant non-zero terms up to an index of 78. To function as a photonic convolution core, we calibrate each MZI stage (Supplementary Information Fig. S7) and apply proportional phase modulations (see Methods). Figure 2e presents the measured circular waveform shifting over time, showing excellent uniformity and smoothness in shifting.

For performance characterization, we perform the convolutional operation across an overlaid FSR between 6250 cm$^{-1}$ and 6670 cm$^{-1}$ (i.e. 1500 nm to 1600 nm), utilizing a benchtop waveshaper (Coherent WS-1000B) in tandem with a broadband source to generate diverse incident spectra. Here, the shifting step $\delta v$ is set at 2.1 cm$^{-1}$ (see inset in Fig. 2e), which corresponds to a sequence length $N$ of 200. This ensures compliance with the Nyquist sampling theorem, meaning that the sampling resolution of the system response sequence $r_{sys}[v]$ must be finer than half the system resolution. Also, this poses a negligible computational load of merely performing Fast Fourier Transforms (FFTs) on a 200-point sequence, which takes about 50 ms running on the MCU (< 0.1ms if executed on a regular laptop).

We begin by examining various discrete, narrowband spectra, including single-, dual-, triple-, and quad-peak signals with different spectral positions and full-width at half-maximum (FWHM). The spectrometer accuracy is quantified by the *L2*-norm relative error $\varepsilon$, defined as $\varepsilon = \|\Phi_0 - \Phi\|_2 / \|\Phi_0\|_2$ where $\Phi$ and $\Phi_0$ denote the resolved and reference spectra, respectively. Figures 2f-h illustrate the precise recovery of these signals, exhibiting low relative errors. The spectral spacing of dual-peak signals is gradually reduced down to 5.8 cm$^{-1}$ (i.e. 1.4 nm), validating a resolution of this value given the Rayleigh criterion. This closely matches the theoretical calculation based on Eq. (6) (i.e. 420 cm$^{-1}$/78 = 5.4 cm$^{-1}$). Subsequently, a range of continuous broadband spectra with randomly shaped waveforms are tested, as shown by Fig. 2i. Our device accurately recovers all complicated spectral features, such as those sharp peaks with rapid roll-offs and minor bumps, achieving ultra-low relative errors between 0.022 and 0.044. To outline the detectability of subtle spectral changes, we create groups of gradually varying waveforms by altering the amplitude or position of specific spectral peaks. Figure 2j presents the resolved waveforms with dipping, rising, and shifting spectral characteristics, respectively, all achieving relative errors lower than 0.035. Notably, a peak intensity as low as 0.02 is identified in the dipping waveforms, which demonstrates a dynamic range of over 17 dB (i.e. in contrast to the maximal power at 1.0).

Thanks to the circular shifting nature, a convolutional spectrometer can readily compensate for the spectral shifts caused by thermal variations, allowing for nearly infinite temperature tolerance. For validation, our device is placed in a temperature chamber, with the ambient temperature changing from −20°C to +80°C. Supplementary Information Section 7 elaborates that this temperature variation only causes a waveform shift of approximately 0.016 nm/°C, owing to the low thermal sensitivity of the SiN platform. Thus, by applying algorithmic correction to counteract the impact of temperature drift, we demonstrate a thermal tolerance of at-least 100°C.

**Spectrometric analysis of solid and liquid substances**

Our ConvSpec empowers the spectrometric analysis of various substances, as shown in Fig. 3a. First, we demonstrate the accurate classification of a diverse range of solid samples, including ten types of commonly used plastics, ten varieties of pharmaceutical powders, ten coffees sourced from different origins, twelve flours with differing gluten content, and eight quality tiers of green tea. Such identifications hold significant importance across areas such as industrial production, sustainable recycling, food safety, and agricultural monitoring. For example, separating recyclable plastics like PET, PVC, and PP enhances their reusability[32], while detecting the gluten content in flour products ensures quality control[33]. Yet, these materials are difficult to distinguish via visual inspection. We repeatedly measure each sample 80 times over the NIR range of 5900 cm$^{-1}$ to 8300 cm$^{-1}$ (i.e. about 1200 nm to 1700 nm) and randomly split the data into training and testing sets in a 2:1 ratio (a partitioning ratio consistently applied across all our spectral modeling tasks). Here, six superluminescent diodes (SLDs) centered at different wavenumbers are employed as light sources to sequentially illuminate the full bandwidth (see Supplementary Information Fig. S9). Figure 3b and 3c plot the measured reflectance spectra for plastic samples and all other samples, with relative errors lower than 0.029, 0.040, 0.028, 0.029 and 0.024, respectively. Notably, our spectrometer exhibits exceptional measurement stability, with the standard deviation of 80 repeated measurements mostly ranging between 0.001 and 0.004 across different samples. Classification models are trained using k-nearest neighbors (kNN) algorithm, all realizing a 100% success rate (Fig. 3d and Supplementary Information Fig. S11).

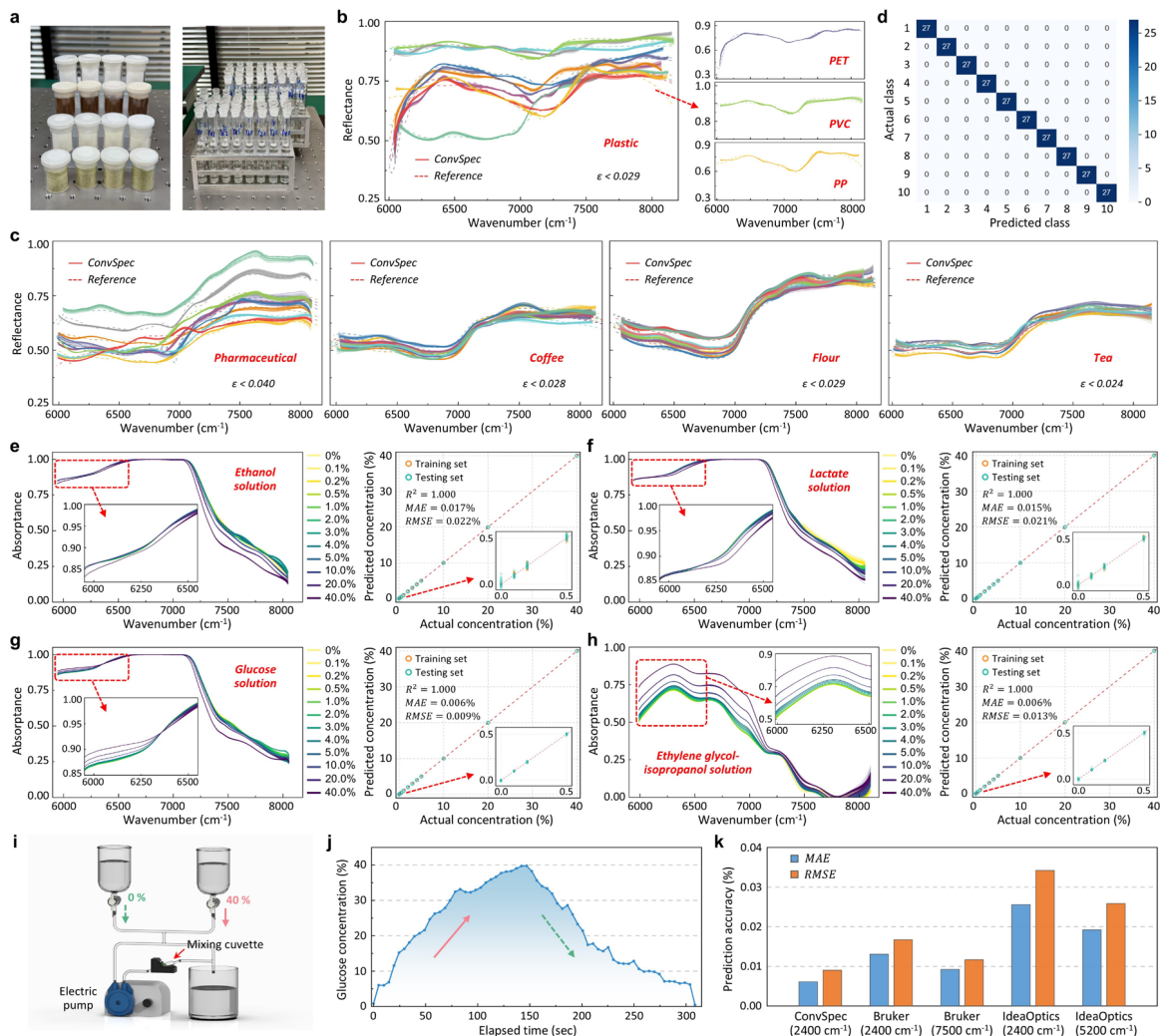

**Fig. 3| Spectrometric measurement of various solid and liquid samples.** (a) Photos of some representative solid and solution samples under test. (b) Measured reflectance spectra of different plastic samples (80 repetitions per sample), with references measured using a benchtop spectrometer. The insets highlight a few representative examples of recyclable plastics. (c) Measured reflectance spectra of different pharmaceuticals, coffee, flour and tea. (d) Confusion matrix of the classification results, using plastic samples as an example. (e-h) Measured absorptance spectra (left) of aqueous solutions (ethanol, lactate, glucose) and organic solutions (EG in IPA) across concentrations from 0.1% to 40% (80 repetitions per concentration), along with the corresponding scatter plots (right) of predicted vs. actual concentrations, all achieving a detection limit of at least 0.1%. (i) Schematic of a dynamic solution mixing system. (j) Real-time tracking of glucose solution concentration using a pre-trained model. Solid and dashed arrows denote the injection of high- and low-concentration solutions, respectively. (k) Comparison of concentration prediction accuracy between our device and commercial benchtop spectrometers in glucose solution tests, using different working bandwidth ranges for data acquisition.

Our spectrometer is also employed for concentration measurements, using aqueous solutions of ethanol, lactate, and glucose, as well as organic solutions of ethylene glycol (EG) in isopropanol (IPA) as representative examples, all prepared in gradient concentrations ranging from 0.1% to 40%. These tests not only validate its ultra-high detection sensitivity, but also highlight the applicability in chemical and pharmaceutical practices, such as detecting EG contamination during manufacturing, which present significant nephrotoxic hazards[34]. Figures 3e-h display the measured absorptance spectra of all solutions (each also repeated for 80 times). Corresponding models are trained using the support vector regression (SVR) algorithm to predict solution concentrations, all achieving a coefficient of determination ($R^2$) of 1.000. Moreover, these models demonstrate exceptional accuracy on their respective testing sets, yielding mean absolute errors (MAE) of 0.017%, 0.015%, 0.006%, and 0.006%, and root mean squared errors (RMSE) of 0.022%, 0.021%, 0.009%, and 0.013%, respectively (see Fig. 3e–h). This indicates a system detection limit of at least 0.1%.

To further highlight the rapid sampling speed of our ConvSpec, we construct a dynamic solution mixing system, as shown by Fig. 3i. Figure 3j illustrates that our device, together with the pre-trained SVR model, effectively tracks the rising and falling of glucose concentrations at second-scale intervals (also see Supplementary Video 1). On the other hand, the glucose concentration tests are also replicated using commercial benchtop spectrometers: a dispersive spectrometer (IdeaOptics NIR17+Px) and a Fourier-transform spectrometer (Bruker MPA II), utilizing their partial and full bandwidths, respectively, both illuminated with a commercial tungsten-halogen lamp source (see Supplementary Information Section 9). Figure 3k illustrates that our spectrometer achieves even lower MAE and RMSE compared to benchtop counterparts, which can be attributed to its measurement stability and noise tolerance.

**Non-invasive sensing of human biomarkers**

Beyond substance analysis, we deploy our ConvSpec in highly dynamic, individual-diverse physiological contexts as a non-invasive sensor for human biomarkers. A wrist wearable probe is developed to serve as the sampling interface, as illustrated in Fig. 4a (also see photos in Supplementary Information Fig. S10). This setup allows a millimeter-scale penetration depth, capturing the reflectance spectra from the epidermis, dermis, and even subcutaneous tissues[35,36]. To establish correlations between complex human spectra and faint biomarker signals, a series of controlled variable experiments are conducted, each varying the levels of skin moisture, blood alcohol, blood lactate, and blood glucose, respectively (experiment procedures detailed in Methods). In total, these experiments involve 126 participant sessions and yield around 6,000 full-band spectra (>1,000 spectra for each biomarker; see statistics in Supplementary Information Section 10). Figure 4b illustrates the optical power losses measured across all participants in their initial experimental states under illumination from six different SLD sources, which range from 16 to 22 dB (i.e. around 1–3% of incident power captured). The inter-individual variations can be attributed to differences in skin tone, body fat percentage, and other physiological factors.

We first examine the sensing of skin moisture. To address the impact of system noise and motion-induced instability on modeling performance, we apply a combination of preprocessing techniques, including outlier removal, baseline drift correction, and moving average smoothing, to the spectral data before feeding it into an SVR model. Figure 4c presents the preprocessed spectra measured from 26 participants, each at varying skin moisture levels. The corresponding multi-participant SVR model achieves a corrected and prediction correlation coefficient (i.e. $R_c$ and $R_p$) of 0.996 and 0.952 on the training and testing sets, respectively, indicating a robust model fitting. Figure 4d plots the predicted vs.

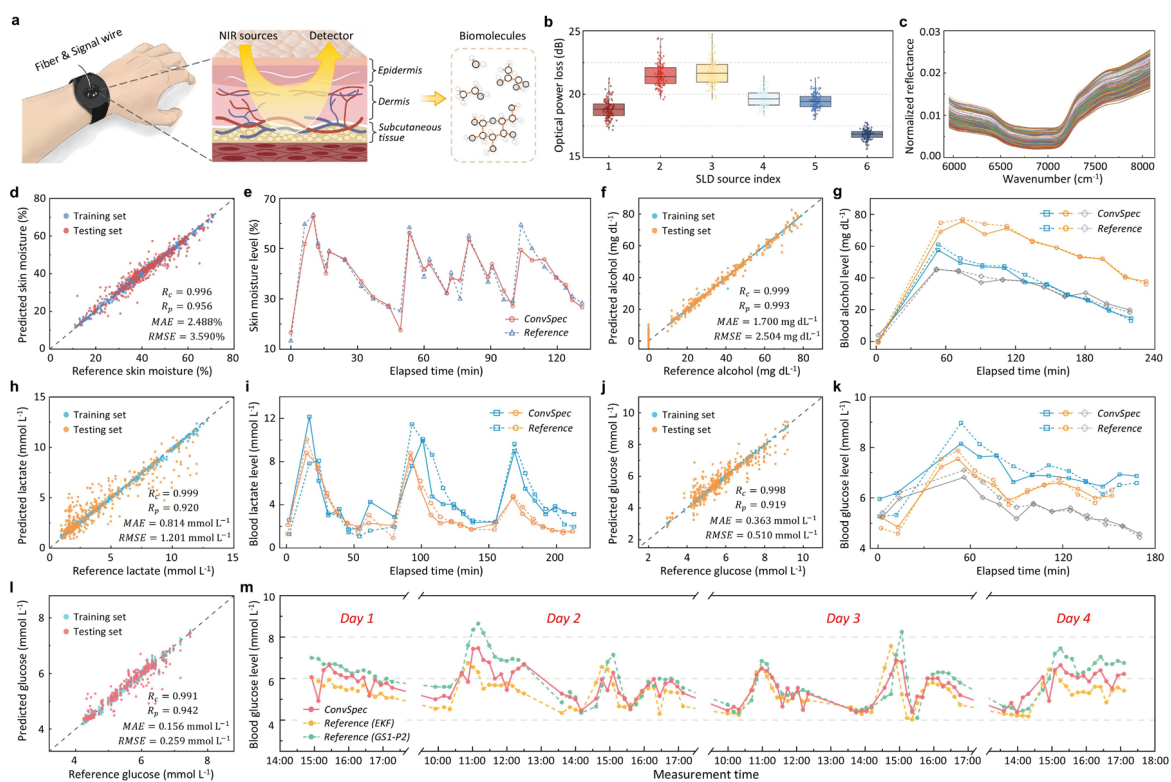

**Fig. 4 | Non-invasive sensing of human biomarkers.** (a) Conceptual diagram of conducting NIR spectroscopic analysis of various biomarkers using our ConvSpec with a wearable probe. (b) Distributions of optical power losses measured from 126 participants under illumination by six different SLD sources. (c) Processed spectra measured at varying skin moisture levels across different participants, normalized to the SLD sources. (d) Predicted vs. reference skin moisture values based on an SVR model. (e) Real-time tracking of a new participant's skin moisture levels across four consecutive rise-and-fall cycles using the pre-trained model. (f,h,j) Predicted vs. reference concentrations of blood alcohol, lactate and glucose based on respective SVR models. (g,i,k) Real-time variations in blood alcohol, lactate, and glucose concentrations from specific participants, comparing with reference values obtained from conventional physicochemical approaches (distinct colors denote different individuals). (l) Predicted vs. reference concentrations of blood glucose based on a single-participant DFNN model. (m) Long-term monitoring of daily glucose concentration fluctuations.

reference values of skin moisture, realizing an MAE of 2.49% and RMSE of 3.59%. Supplementary Information Fig. S13a further illustrates the distribution of prediction errors, showing an 80th percentile at 3.98%. To validate the model's transferability, additional measurements are performed on a new participant undergoing four consecutive cycles of alternating skin moisture levels (i.e. whose spectral data were unseen during the initial model training). As shown in Fig. 4e, the pre-trained model accurately extracts moisture fluctuations between about 10% to 60%, delivering an MAE of 2.61% and an RMSE of 3.62%.

We further investigate the subsurface detection of critical blood metabolites, which is particularly challenging due to their low concentrations and subtle molecular signatures[37,38]. As for blood alcohol, we measure spectra from 24 participants throughout their individual progression from sobriety to mild intoxication, as shown in Supplementary Information Fig. S14b. An SVR model is then trained to predict alcohol levels, achieving an MAE of 1.70 mg dL$^{-1}$ and an RMSE of 2.50 mg dL$^{-1}$ (Fig. 4f). For example, Fig. 4g showcases the real-time blood alcohol variations from three representative participants, exhibiting peak values between 60–80 mg dL$^{-1}$ that gradually decline to 20–40 mg dL$^{-1}$ over three hours. Likewise, blood lactate sensing is assessed by collecting spectra from 48 participants undergoing intense anaerobic exercise (see Supplementary Information Fig. S14c). Figure 4h plots the modeling results, with an MAE and RMSE of 0.814 mmol L$^{-1}$ and 1.201 mmol L$^{-1}$, respectively. Figure 4i presents the temporal variations in lactate concentration from two specific participants, showing distinct peaks and valleys that correspond to three consecutive exercise sessions. Notably, the modeling accuracy for blood lactate can be further enhanced by incorporating heart rates as an additional data dimension, reducing the MAE and RMSE to 0.454 mmol L$^{-1}$ and 0.669 mmol L$^{-1}$, respectively (see Supplementary Information Section 12).

Importantly, our ConvSpec demonstrates its potential in non-invasive blood glucose monitoring—a prominent pursuit in the field that, however, has seen limited breakthroughs over the past decades[39–41]. We record the spectra from 27 participants undergoing a standard oral glucose tolerance test (OGTT; see methods)[13], as shown in Supplementary Information Fig. S14d. The resulting SVR model achieves an MAE and RMSE of 0.363 mmol L$^{-1}$ and 0.510 mmol L$^{-1}$, respectively (Fig. 4j). As examples, Fig. 4k depicts three participants' blood glucose variations during the OGTT, respectively. Moreover, we perform a long-term trial to monitor daily glucose fluctuations on a particular participant. This task is profoundly difficult due to not only the limited fluctuation range caused by daily dietary habits, but also the significant physiological changes over time, which may easily obscure glucose-related spectral features. To tackle this, we develop a personalized model based on a deep feedforward neural network (DFNN) to enhance feature extraction (see Methods). Figure 4l shows the modeling performance, realizing an MAE and RMSE of 0.156 mmol L$^{-1}$ and 0.259 mmol L$^{-1}$, respectively. Figure 4m presents the retrieved glucose fluctuations over multiple days, realizing mean absolute relative differences (MARD) of 7.82% and 10.03% against the references, respectively. Given the scale of dataset collected so far, neither the trained multi-participant nor single-participant glucose model can be directly transferred to new participants, as inter-individual variability compromises the model correlation (see discussions in Supplementary Section 13). This limitation has to be addressed by largely expanding the database.

## Discussion and Conclusion

Table 1 summarizes the characteristics of the proposed convolutional spectrometer against other types, highlighting its theoretical performance advantages. Also, we outline the performance metrics of state-of-the-art miniaturized spectrometers based on various working principles and material platforms, as listed in Supplementary Table S3. By comparison, our device not only excels in key metrics such as bandwidth, precision, and sampling speed, but also stands out in terms of technical readiness levels and fabrication cost. In addition, Supplementary Information Table S4 reviews a variety of NIR spectroscopic studies targeted for different application scenarios, including those utilizing commercial portable and benchtop spectrometers, as well as prototype-level miniaturized sensors. The results illustrate that our spectrometer delivers competitive modeling correlation and accuracy to that of commercial counterparts.

**Table 1. Performance characteristics of different spectrometer categories**

| Principle | Bandwidth | Resolution | Computation complexity | System noise | Detection channel | Fellgett's advantage |
|---|---|---|---|---|---|---|
| Dispersive/ Narrowband filtering (spatial) | D.D. | Filter's FWHM | $O(N)$ | $\sigma$ | $N$ | No |
| Narrowband filtering (temporal) | D.D. | Filter's FWHM | $O(N)$ | $\sigma$ | 1 | No |
| Fourier transform (spatial) | $\frac{1}{OPD_{max}}$ | $\frac{1}{2\delta OPD}$ | $O(Nlog(N))$ | $\sqrt{N}\sigma$ | $\frac{OPD_{max}}{\delta OPD}$ | No |
| Fourier transform (temporal) | $\frac{1}{OPD_{max}}$ | $\frac{1}{2\delta OPD}$ | $O(Nlog(N))$ | $\sqrt{N}\sigma$ | 1 | Yes |
| Computational reconstruction | D.D. | D.D. | $O(N^3)$ | Nonlinear distortion | D.D. | D.D. |
| This work: Convolutional spectrometer | unlimited ($FSR_{comp}$ per operation) | $\frac{FSR_{comp}}{max\{k|R_{sys}[k]>0\}}$ | $O(Nlog(N))$ | $\sigma$ | 1 | Yes |

\* D.D.: design dependent; FWHM: full width at half maximum; $\sigma$: standard deviation of the measurement noise; $N$: number of spectral pixels; $OPD$: optical path difference; $FSR_{comp}$ and $R_{sys}[k]$ denote the composite FSR and the frequency component sequence of system response, respectively, respectively (see Eq. (1-6)).

In summary, we define a new class of spectrometer based on the convolution theorem and leverage it to realize a variety of NIR applications from chemical analysis to biomedical sensing. Our work represents the transformation of miniaturized spectrometers towards low-cost, high-performance, and embeddable spectrometric sensors, offering wide application potentials in industrial manufacturing, food quality control, and environmental monitoring. Moreover, the successful demonstration of non-invasive biomarker sensing opens practical opportunities for next-generation healthcare solutions, enabling intelligent systems for hydration assessment, intoxication alerts, fitness tracking, and real-time glucose monitoring in diabetes management.

## Method

### Circular waveform shifting in the spectral domain

For an unbalanced MZI $i$, its periodic response is written as[42]:

$$r_i(\nu, t) = \rho^2 + (1-\rho)^2 + 2\rho(1-\rho)\cos(2\pi n_{eff}\nu\Delta L_i + \varphi_i(t)) \quad (7)$$

where $\rho$ is the power splitting ratio of directional couplers. The oscillation term $\cos(2\pi n_{eff}\nu\Delta L_i + \varphi_i(t))$ is dictated by both arm length difference $\Delta L_i$ and temporal phase shift $\varphi_i(t)$. Accordingly, the overlaid system response can be denoted as:

$$r_{sys}(\nu, t) = \prod_{i=1}^{n_{stage}}(\gamma_1(\gamma_2 + \cos(2\pi n_{eff}\nu\Delta L_i + \varphi_i(t)))) \quad (8)$$

where $\gamma_1 = 2\rho(1-\rho)$ and $\gamma_2 = (\rho^2 + (1-\rho)^2)/2\rho(1-\rho)$. As shown by Fig. 1c, the time-varying phase shift at each MZI $\varphi_i(t)$ introduces a circular waveform shift within its FSR. Likewise, when the variations in phase shift across different MZI stages are proportional to their respective arm length differences, i.e. adhering to the following equation:

$$\mathrm{mod}\{\Delta\varphi_1(t):\Delta\varphi_2(t):\cdots:\Delta\varphi_N(t), 2\pi\} = \frac{1}{FSR_1}:\frac{1}{FSR_2}:\cdots:\frac{1}{FSR_i} = \Delta L_1:\Delta L_2:\cdots:\Delta L_i \quad (9)$$

where $\mathrm{mod}\{\ldots, 2\pi\}$ represents the modulo operation of phase shifts with respect to $2\pi$, the overlaid waveform can be circularly shifted over the composite FSR. This process can be described by reformulating Eq. (8) in terms of finite differences, as:

$$\begin{aligned} r_{sys}(\nu, t_0 + \Delta t) &= \prod_{i=1}^{n_{stage}}\left(\gamma_1\left(\gamma_2 + \cos\left(2\pi n_{eff}\nu\Delta L_i + \varphi_i(t_0 + \Delta t)\right)\right)\right) \\ &= \prod_{i=1}^{n_{stage}}\left(\gamma_1\left(\gamma_2 + \cos\left(2\pi n_{eff}(\nu + \Delta\nu)\Delta L_i + \varphi_i(t_0)\right)\right)\right) \\ &= r_{sys}(\nu + \Delta\nu, t_0) \end{aligned} \quad (10)$$

where $\Delta\nu = \frac{\Delta\varphi_i(\Delta t)}{2\pi n_{eff}\Delta L_i}$. Eq. (10) elucidates how the convolutional core converts time-domain phase modulation into a spectral-domain waveform shifting.

### Chip fabrication and packaging

The spectrometer chip was fabricated on the CORNERSTONE SiN integration platform, which employs standard deep ultraviolet (DUV) lithography with a 250 nm feature size. The platform comprises a 300 nm thick SiN layer sandwiched between a 3 μm buried oxide layer and a 2 μm silicon dioxide top cladding layer. The chip was wire-bonded to a customized PCB for electrical fan-out and edge-coupled to lensed polarization maintaining fibers (PMFs) for optical assessment (see Fig. 2a). UV-curable adhesive was applied to mechanically secure the PMFs, ensuring a coupling loss of around 2.5 dB per facet. A thermoelectric cooler was placed underneath the chip, which operated in tandem with a thermistor to realize temperature stabilization.

**PCB design**

A compact PCB was developed to facilitate full packaging of the spectrometer chip, as shown in Fig. 2b. The PCB integrated an MCU (STM32), a fiber-coupled InGaAs PD (LSIPD-L0.3) with transimpedance amplifier (TIA) read-out circuit, driving circuits based on 8-channel digital-to-analog converters (DACs) and operational amplifiers, as well as the associated power management module and data communication port. The MCU controlled the driving voltages and calculated the convolution operation for spectra recovery, achieving an overall sampling and processing time of less than 0.4 seconds per spectrum.

**Optical testbed and sampling schemes**

To calibrate the system spectral response (Fig. 2c), we sequentially launched six SLDs located at different wavenumbers as broadband light sources and measured the corresponding output spectra via a commercial benchtop spectrum analyzer. The obtained full-band system response was then stored in the MCU. During the spectroscopic sensing of various substances and human biomarkers, these SLDs were also sequentially turned on for sample illumination. To accommodate their bandwidths, we utilized another chip variant with the overlaid FSR doubled to 840 cm$^{-1}$. This enables the convolutional spectral recovery within each SLD's coverage range, thereby forming the sample's full-band reflectance/absorbance spectra. The schematic of the testbed workflow and the ASE spectra of the SLDs are shown in Supplementary Information Fig. S9.

For fiber-confined signals (e.g. those in Fig. 2), measurements were taken using the fiber-coupled PD on the PCB. Meanwhile, different free-space sampling interfaces were developed to facilitate our spectrometric experiments. Supplementary Information Figs. S10a-b detail the sampling schemes for measuring the reflectance and absorptance of solid and liquid samples, respectively, where a single-mode fiber collimator was used to launch light onto the samples, and PDs received the reflected or transmitted signals. The resulting spectra were then normalized with standard references, i.e. a fully reflective whiteboard or an empty cuvette. As for the biomarker sensing, a wrist wearable probe was customized to measure the reflection spectra from human skin. Supplementary Information Fig. S10c illustrates the schematic of this probe, featuring a centrally positioned fiber collimator surrounded by PDs to capture the reflected signal from different skin depths.

**Preparation of solid and liquid samples**

All solid samples used in our experiments are common industrial, medical, and agricultural products. Specifically, the ten plastics include polyethylene terephthalate (PET), polyvinyl chloride (PVC), polypropylene (PP), polycarbonate (PC), polyvinyl alcohol (PVA), acrylonitrile styrene (AS), acrylonitrile butadiene styrene (ABS), polytetrafluoroethylene (PTFE), polybutylene adipate terephthalate (PBAT), and polyhydroxybutyrate (PHB). The ten pharmaceuticals comprise ibuprofen, paracetamol, povidone-iodine, cefaclor, diclofenac sodium, digestive enzymes, roxithromycin, inosine, riboflavin (Vitamin B2), and glimepiride. Coffee samples are produced from various regions, such as Yunnan, Colombia, Rwanda, and East Java, which also differ in tree species and roasting methods. Flour samples span a range of gluten contents, from low-gluten to high-gluten grades, while the green tea samples vary in quality, ranging from standard to premium grades. These samples are ground and sieved into uniformly sized particles and then placed into glass containers for reflectance measurements.

The liquid samples are all carefully prepared at varying concentrations, where the glucose solutions are measured by mass percentage, while the ethanol, lactate, and EG solutions are prepared by volume percentage. These solutions are then transferred into cuvettes for absorptance measurements.

**Data collection procedure for biomarker analysis**

To establish datasets correlating human spectra with various biomarkers, we utilized our ConvSpec for spectral collection while simultaneously employing different physicochemical methods to obtain corresponding reference values from participants. Throughout all biomarker experiments, the skin reflectance spectra were repetitively measured at the same area on the wrist using our wearable probe. Participants were required to keep their forearms relaxed, the skin clean, and the probe evenly pressed against skin surface without excessive pressure or loose gaps. Meanwhile, the temperature and humidity of the testing environment were strictly controlled to ensure data stability and consistency.

During the testing of skin moisture, a commercial conductivity-based moisture meter was used to record the real-time moisture percentage. We first employed an air dryer to reduce skin moisture to below 20% as the starting point for data collection. Afterward, a measured amount of moisturizer was applied to participants' skin and allowed to absorb for 5 minutes before being wiped off. Skin moisture and reflectance spectra were then measured simultaneously. This measurement process was repeated every 3 minutes until no further increase in moisture was observed.

For the blood alcohol testing, a commercial breathalyzer served as reference. All participants consumed over 200 ml of high-proof baijiu (42% ABV) in a semi-satiated state, which ensures peak blood alcohol concentrations of at least 40 mg dL$^{-1}$. Breath alcohol tests were conducted at 15-minute intervals from before drinking until three hours afterward, while skin reflectance spectra were simultaneously recorded using our ConvSpec.

As for blood lactate and glucose, reference values were obtained via fingertip blood samples, analyzed using a benchtop lactate-glucose analyzer (EKF Biosen C-Line Clinic) via an electrochemical approach. Specifically, in blood lactate experiments, participants first engaged in 60 to 90 seconds of intense cycling, which typically elevates the blood lactate levels to around 10 mmol L$^{-1}$. Immediately after exercise, fingertip blood samples were taken, while the skin reflectance spectra and heart rate data were also recorded. Such data collection was repeated every 8 minutes until the blood lactate levels dropped to below 2.5 mmol L$^{-1}$ (around 48 minutes after exercise). Each participant completed two to three consecutive sessions of such exercise-rest alternation. While in blood glucose experiments, participants engaged in a standard oral glucose tolerance test by consuming 75g of anhydrous glucose in a fasting state. This led to a rise in their blood glucose levels, typically peaking at about 8 mmol/L, followed by a gradual decline. Fingertip blood samples and skin reflectance spectra were collected at 10-minute intervals from before the glucose consumption to 3 hours afterward.

During the experiment of long-term blood glucose tracking, the particular participant not only performed the fingertip blood glucose measurements, but also wore an invasive glucose probe (GS1-P2, Sibionics) as an additional reference for continuous glucose monitoring.

**Statistical analysis of participants and ethical approval**

To ensure sample diversity and data richness, we recruited male and female Asian adult participants across various age groups for the biomarker sensing experiments. Supplementary Information Table S2 details the number of participants, gender ratio, and age distribution among various biomarker sensing experiments.

All procedures involving human participants were conducted in accordance with the 1964 Helsinki Declaration and its later amendments or comparable ethical standards. Appropriate ethics approval and informed written consent from all participants were obtained. The experiments involving fingertip blood collection were performed by experienced professionals, ensuring hygiene and the health of participants throughout the collection process.

**Data processing and modeling**

For the spectrometric analysis of solid samples, we develop classification models using the kNN algorithm, which determines the majority class among the k-nearest neighbors in the feature space. We iteratively adjust the value of k (the number of nearest neighbors) to realize maximal classification accuracy on the training set. The refined model is then evaluated on the testing set. As for predicting solution concentrations, we employ the SVR algorithm, a supervised machine learning technique that maps input data into a high-dimensional feature space and constructs an optimal hyperplane to model nonlinear relationships. We fine-tune the SVR model's regularization and kernel parameters to optimize the coefficient of determination ($R^2$) and RMSE on the training set, thereby enabling accurate prediction in the testing set.

During the multi-participant modeling of various biomarkers, we first apply a series of preprocessing steps to improve data quality. The processed data then undergoes the sample set partitioning based on joint X-Y Distances (SPXY) algorithm to ensure uniform distribution between the training and testing sets. Corresponding SVR models are trained, with the radial basis function (RBF) chosen as the kernel for its effectiveness in capturing nonlinear dependencies. Key parameters are fine-tuned using $R^2$ and RMSE as optimization metrics. For the single-participant modeling of long-term glucose monitoring, we also we preprocess the data before inputting it into a DFNN, which consists of multiple layers of feature extraction and regression sections. In specific, the feature extraction layers process the input through layers with 512, 256, and 128 nodes, each equipped with batch normalization, LeakyReLU activation, and Dropout to enhance stability and prevent overfitting. The regression layers further refine the extracted features using layers 128, 64, and 32 nodes, and ultimately produce a single output for glucose concentration prediction. The model completes over 3000 training iterations to ensure prediction accuracy.

In addition, to verify the independence of our models, cross-analysis is performed by transferring data from one model into another. For instance, Supplementary Information Fig. S17 confirms the independence between the models for blood glucose and lactate. Meanwhile, we evaluate the modeling performance across different algorithms, as detailed in Supplementary Information Section 16.

**Data availability**

All data supporting this study are included within the main text and/or Supplementary Information. Source data are provided with this paper. Additional inquiries regarding the data can be directed to the corresponding author.

**Code availability**

A series of demo codes are available from the public repository at GitHub (https://github.com/GlitterinTechRnD/ConvSpec), which include the convolutional recovery of unknown spectra, the kNN and SVR models for solid and liquid substances, and the SVR and DFNN models for biomarker analysis.

**Reference**

1. Bao, J. & Bawendi, M. G. A colloidal quantum dot spectrometer. *Nature* **523**, 67–70 (2015).
2. Yang, Z. *et al.* Single-nanowire spectrometers. *Science* **365**, 1017–1020 (2019).
3. Yoon, H. H. *et al.* Miniaturized spectrometers with a tunable van der Waals junction. *Science* **378**, 296–299 (2022).
4. Crocombe, R. A. Portable spectroscopy. *Applied Spectroscopy* **72**, 1701–1751 (2018).

5. Yang, Z., Albrow-Owen, T., Cai, W. & Hasan, T. Miniaturization of optical spectrometers. *Science* **371**, eabe0722 (2021).
6. Parker, F. *Applications of Infrared Spectroscopy in Biochemistry, Biology, and Medicine*. (Springer Science & Business Media, 2012).
7. Pasquini, C. Near infrared spectroscopy: A mature analytical technique with new perspectives – A review. *Analytica Chimica Acta* **1026**, 8–36 (2018).
8. Bacon, C. P., Mattley, Y. & DeFrece, R. Miniature spectroscopic instrumentation: applications to biology and chemistry. *Review of Scientific instruments* **75**, 1–16 (2004).
9. Baker, M. J. *et al.* Using Fourier transform IR spectroscopy to analyze biological materials. *Nat Protoc* **9**, 1771–1791 (2014).
10. Beć, K. B., Grabska, J. & Huck, C. W. Near-Infrared Spectroscopy in Bio-Applications. *Molecules* **25**, 2948 (2020).
11. Chitnis, D. *et al.* Towards a wearable near infrared spectroscopic probe for monitoring concentrations of multiple chromophores in biological tissue in vivo. *Review of Scientific Instruments* **87**, 065112 (2016).
12. Martin, F. L. *et al.* Distinguishing cell types or populations based on the computational analysis of their infrared spectra. *Nat Protoc* **5**, 1748–1760 (2010).
13. Heise, H. M., Delbeck, S. & Marbach, R. Noninvasive Monitoring of Glucose Using Near-Infrared Reflection Spectroscopy of Skin—Constraints and Effective Novel Strategy in Multivariate Calibration. *Biosensors* **11**, 64 (2021).
14. Lafrance, D., Lands, L. C. & Burns, D. H. Measurement of lactate in whole human blood with near-infrared transmission spectroscopy. *Talanta* **60**, 635–641 (2003).
15. Yao, C. *et al.* Chip-scale sensor for spectroscopic metrology. *Nat Commun* **15**, 10305 (2024).
16. Li, A. *et al.* Advances in cost-effective integrated spectrometers. *Light Sci Appl* **11**, 174 (2022).
17. Shekhar, S. *et al.* Roadmapping the next generation of silicon photonics. *Nat Commun* **15**, 751 (2024).
18. Redding, B. Compact spectrometer based on a disordered photonic chip. *Nat Photonics* **7**, 6 (2013).
19. Kita, D. M. *et al.* High-performance and scalable on-chip digital Fourier transform spectroscopy. *Nat Commun* **9**, 4405 (2018).
20. Yao, C. *et al.* Integrated reconstructive spectrometer with programmable photonic circuits. *Nat Commun* **14**, 6376 (2023).
21. Cheben, P. *et al.* A high-resolution silicon-on-insulator arrayed waveguide grating microspectrometer with sub-micrometer aperture waveguides. *Optics express* **15**, 2299–2306 (2007).
22. Zhang, L. *et al.* Ultrahigh-resolution on-chip spectrometer with silicon photonic resonators. *Opto-Electronic Advances* **5**, 210100–1 (2022).
23. Montesinos-Ballester, M. *et al.* On-chip Fourier-transform spectrometer based on spatial heterodyning tuned by thermo-optic effect. *Sci Rep* **9**, 14633 (2019).
24. Guan, Q., Lim, Z. H., Sun, H., Chew, J. X. Y. & Zhou, G. Review of Miniaturized Computational Spectrometers. *Sensors* **23**, 8768 (2023).
25. Yao, C. *et al.* Benchmarking Reconstructive Spectrometer with Multiresonant Cavities. *ACS Photonics* (2024) doi:10.1021/acsphotonics.4c00915.
26. Xu, H., Qin, Y., Hu, G. & Tsang, H. K. Cavity-enhanced scalable integrated temporal random-speckle spectrometry. *Optica,* **10**, 1177–1188 (2023).
27. Wan, Y., Fan, X. & He, Z. Review on Speckle-Based Spectrum Analyzer. *Photonic Sens* **11**, 187–202 (2021).
28. Yao, C. *et al.* Broadband picometer-scale resolution on-chip spectrometer with reconfigurable photonics. *Light Sci Appl* **12**, 156 (2023).
29. Chen, C., Gu, H. & Liu, S. Ultra-simplified diffraction-based computational spectrometer. *Light Sci*

*Appl* **13**, 9 (2024).
30. Krishna, H. *Digital Signal Processing Algorithms: Number Theory, Convolution, Fast Fourier Transforms, and Applications*. (Routledge, New York, 2017).
31. Vaseghi, S. V. *Advanced Digital Signal Processing and Noise Reduction*. (John Wiley & Sons, 2008).
32. Alassali, A., Fiore, S. & Kuchta, K. Assessment of plastic waste materials degradation through near infrared spectroscopy. *Waste Management* **82**, 71–81 (2018).
33. Sun, D.-W. *Infrared Spectroscopy for Food Quality Analysis and Control*. (Academic Press, 2009).
34. Bodo, E., Merlo, S. & Bello, V. Spectral Fingerprint Investigation in the near Infra-Red to Distinguish Harmful Ethylene Glycol from Isopropanol in a Microchannel. *Sensors* **22**, 459 (2022).
35. Alsunaidi, B., Althobaiti, M., Tamal, M., Albaker, W. & Al-Naib, I. A Review of Non-Invasive Optical Systems for Continuous Blood Glucose Monitoring. *Sensors* **21**, 6820 (2021).
36. Ge, Q. *et al.* Evaluation and Validation on Sensitivity of Near-Infrared Diffuse Reflectance in Non-Invasive Human Blood Glucose Measurement. *Sensors* **24**, 5879 (2024).
37. Heise, H. M. Non-Invasive Monitoring of Metabolites Using Near Infrared Spectroscopy: State of the Art. *Hormone and Metabolic Research* **28**, 527–534 (2007).
38. Huber, M. *et al.* Stability of person-specific blood-based infrared molecular fingerprints opens up prospects for health monitoring. *Nat Commun* **12**, 1511 (2021).
39. Khalil, O. S. Non-Invasive Glucose Measurement Technologies: An Update from 1999 to the Dawn of the New Millennium. *Diabetes Technology & Therapeutics* **6**, 660–697 (2004).
40. Villena Gonzales, W., Mobashsher, A. T. & Abbosh, A. The Progress of Glucose Monitoring—A Review of Invasive to Minimally and Non-Invasive Techniques, Devices and Sensors. *Sensors* **19**, 800 (2019).
41. Tang, L., Chang, S. J., Chen, C.-J. & Liu, J.-T. Non-Invasive Blood Glucose Monitoring Technology: A Review. *Sensors* **20**, 6925 (2020).
42. Saghaei, H., Elyasi, P. & Karimzadeh, R. Design, fabrication, and characterization of Mach–Zehnder interferometers. *Photonics and Nanostructures - Fundamentals and Applications* **37**, 100733 (2019).

**Acknowledgement**

This research was mainly supported by GlitterinTech Limited, but also received support from the UK EPSRC through project QUDOS (EP/T028475/1) and European Union's Horizon 2020 Research and Innovation program through project INSPIRE (101017088). The authors thank Mr. Bobo Liu, Ms. Mengting Wu, Ms. Yuting Wen, Mr. Jinbin Wang, Mr. Yanlong Liang, Mr. Keming Qiu, and Mr. Yahang Chen for their assistance in experiments. The authors also thank Ms. Jinglei Cao, and Ms. Lanjun Xu for contributions to figure preparation.

**Author contributions**

C.Y. and J.M. jointly conceived the concept of the convolutional spectrometer. C.Y. developed the theoretical framework, conducted the optical simulations (with assistance from N.W.), and designed the photonic chip. J.M. implemented the chip packaging and designed the associated circuitry and sampling interfaces. C.Y. conducted the spectroscopic sensing experiments for substances and human biomarkers, with assistance from W. Zhuo, Z.T., J.M., L.M., and T.Y. C.Y. drafted the manuscript, with input from N.W., P.B., W. Zhang, K.X., Y.Y., T.H., I.W., R.P., and Q.C. The project was supervised by R.P. and Q.C.

**Competing interests**

GlitterinTech Limited declares a pending patent application filed with the China National Intellectual Property Administration (inventors: C.Y. and J.M., application number: CN202410920608X), which relates to the spectrometer design presented in this manuscript. This patent was also pursued as a pending PCT application (application number: PCT/CN2024/109952). The authors declare no other competing interests.